\documentclass [prb,aps,letterpaper,twocolumn,amsmath,amssymb,floatfix] {revtex4}

\usepackage{graphicx}
\usepackage{textcomp}
\usepackage{mathptmx} 

\newcommand{\dgc}{$^\circ\mathrm{C}$}

\begin{document}

\title{Electric-field-dependent spectroscopy of charge motion using a
single-electron transistor}

\author{K. R. Brown}
\email{kbrown@lps.umd.edu}
\author{L. Sun}
\author{B. E. Kane}
\affiliation{Laboratory for Physical Sciences, 8050 Greenmead Drive,
  College Park, Maryland, 20740}


\begin{abstract}
We present observations of background charge fluctuators near an
Al-AlO$_x$-Al single-electron transistor on an oxidized Si
substrate. The transistor design incorporates a heavily doped
substrate and top gate, which allow for independent control of the
substrate and transistor island potentials. Through controlled
charging of the Si/SiO$_2$ interface we show that the fluctuators
cannot reside in the Si layer or in the tunnel barriers. Combined with
the large measured signal amplitude, this implies that the defects
must be located very near the oxide surface.
\end{abstract}


\maketitle

Metal single-electron transistors (SETs) have generated interest
recently for possible applications in quantum information processing,
\cite{Kane2} nanoelectronics,\cite{Sunamura1} and fundamental
sensing.\cite{LaHaye1} They are known to be sensitive electrometers
with a charge noise around
$10^{-6}$~$e/\sqrt{\mathrm{Hz}}$.\cite{Aassime1} Nevertheless,
SETs have not yet fulfilled their promise in
many areas due to their broad $1/f^\alpha$ noise spectrum and large
background charge fluctuations. Identification and elimination of the
sources of these nonidealities will be important for the fabrication
of practical SET devices.

Several groups have attempted to isolate the source of noise in metal
SETs.\cite{Furlan1,Buehler1,Zorin1,Krupenin1,
Tavkhelidze1,Zimmerman1,Kenyon1} Furlan and Lotkhov used a
configuration with four surface gates to locate charge traps very near
the island of an Al SET.\cite{Furlan1} Buehler \emph{et al}.\ used a
two-gate geometry to study two-level fluctuations in their device and
identified sources both near the surface and in the bulk
substrate.\cite{Buehler1} However, there is as yet no consensus for
the dominant noise source in metal SETs. Of particular interest are
fluctuators that become evident when large electric fields ($\gtrsim
1$~kV/cm) are applied to an SET. This regime is particularly relevant
for silicon quantum computing proposals\cite{Kane1,Kane2} that rely
on an SET for final state determination, for displacement detection
with an SET,\cite{LaHaye1, Knobel1} and for future SET-based
electronic devices such as memories.\cite{Chen1, Sunamura1}

We have identified similar irregularities in Al-AlO$_x$-Al SETs
fabricated on oxidized Si substrates. These defects manifest
themselves as peaks in the polarizability of the gate-island
dielectric as a function of gate potential. They are present in every
SET device we have measured to date. Through the use of a multiple
gate geometry we have determined that the defects associated with
these irregularities must lie above the substrate surface. Thus, they
would seem to be a typical feature of Al-AlO$_x$-Al SETs regardless
of substrate.

\begin{figure}[tb!]
  \centering \includegraphics*{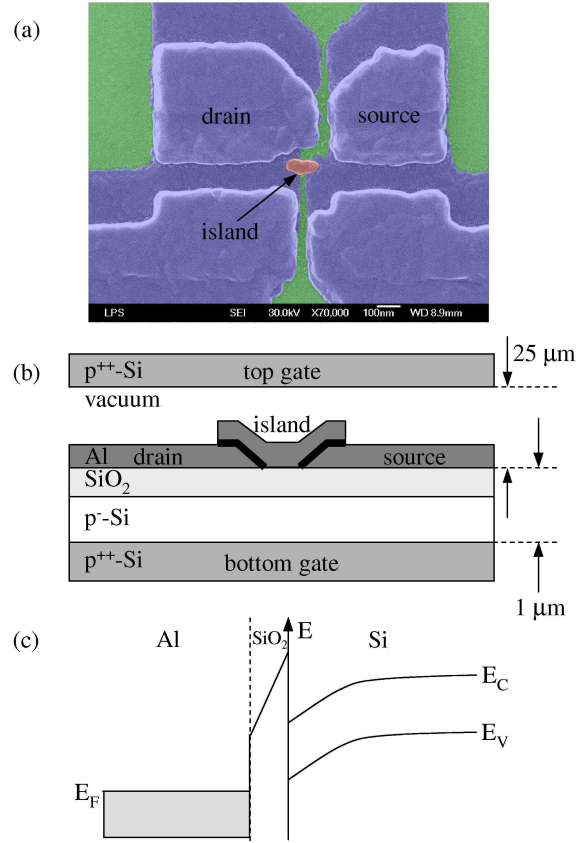}
\caption{(a) Colorized SEM micrograph of a typical device. Green areas
  represent the substrate, blue areas the SET leads, and the red area
  the SET island. The island dimensions are 60~nm$\times$130~nm,
  although only about half of this area is in contact with the
  substrate due to overlap with the leads.  The leads extend farther
  than 600~nm from the island in any direction. (b) Cross sectional
  schematic of the SET geometry. The substrate is Si with 20~nm of
  surface thermal oxide. The Al source, island, and drain of the SET
  lie on the wafer surface. The heavily p-doped bottom gate has a peak
  density 1~$\mu$m below the surface. The top gate is suspended
  25~$\mu$m above the chip. (c) Band diagram of the device in
  inversion. The electric field drives electrons toward the interface
  (left) and holes toward the substrate (right).}
\label{fig:set geometry}
\end{figure}

Each of our devices consists of an Al-AlO$_x$-Al single-electron
transistor on an oxidized Si substrate. Figure\ \ref{fig:set geometry}
shows a scanning electron microscopy (SEM) micrograph of one device, a
cross sectional schematic, and a band diagram of the device geometry.
We intentionally make the leads of the SET wide so that the
system is effectively three parallel plates, as shown schematically in
Fig.\ \ref{fig:set geometry}(b), thereby making the electrostatics of
the device easier to model. Electric fields from the top and bottom
gates are confined for the most part above and below the SET,
respectively.

Device fabrication begins with a nearly intrinsic Si wafer ($\rho >
10$~k$\Omega$-cm). The wafer is implanted with B at an energy of
500~keV and an areal density of 2.5$\times$10$^{14}$~/cm$^2$ to form
the bottom gate. The peak acceptor density of
6$\times$10$^{18}$~/cm$^3$, high enough to conduct at low
temperatures, occurs 1~$\mu$m below the surface. The wafer is then
oxidized in a tube furnace at 950 \dgc\ to a thickness of 20~nm. We
use electron-beam lithography and self-aligned double-angle
evaporation to make the SETs.\cite{Fulton1} The contact area between
the SET island and substrate is of the order 50$\times$50~nm$^2$. The
top gate consists of a piece of heavily doped Si, glued by hand to the
wafer surface, but separated from it by 25~$\mu$m Kapton spacers at
either end. The devices are cooled below 100~mK in an Oxford dilution
refrigerator. A 1~T magnetic field is applied to keep them in the
normal state.

\begin{figure}
  \centering \includegraphics*{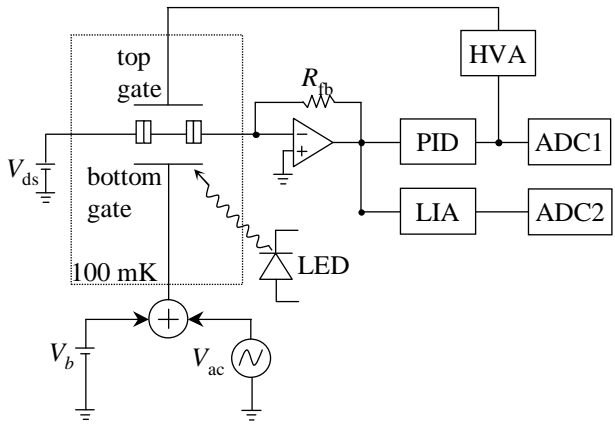}
\caption{Schematic of the measurement setup. LIA, lock-in amplifier;
  PID, proportional-integral-differential feedback; HVA, high-voltage
  amplifier; and ADC, analog-to-digital converter, respectively.}
\label{fig:pid lockin schematic}
\end{figure}

Figure\ \ref{fig:pid lockin schematic} shows a schematic of the
measurement setup.
For fixed drain-source bias $V_\mathrm{ds}$, the current
$I_\mathrm{ds}$ flowing through the device is monitored via a
room-temperature current preamplifier with a 3~kHz bandwidth. As the
potential $V_b$ on the bottom gate is changed, a
proportional-integral-differential (PID) feedback circuit maintains a
constant $I_\mathrm{ds}$ by controlling the top gate voltage $V_t$,
ensuring that the SET remains biased for maximum sensitivity. The
feedback has a bandwidth of 10~Hz or less. To induce another electron
onto the SET island using the top gate requires a voltage change
around 60~V, so that a high-voltage amplifier (HVA) is required to
amplify the output of the PID circuit into a useful range. A small ac
dither $V_\mathrm{ac}$ at 1~kHz is applied simultaneously with $V_b$
to the bottom gate, and the response of the SET at 1~kHz is measured
with a lock-in amplifier (LIA). This signal is well above the 10~Hz
bandwidth of the PID circuit, and so is unaffected by the latter. Both
the PID output voltage, which reflects changes in the dc island
potential, and the in-phase ac response (lock-in X channel) are then
recorded while sweeping $V_b$. The setup includes an optical fiber to
illuminate the sample with a light-emitting diode (LED) mounted at
room temperature near the top of the refrigerator.

\begin{figure}
  \centering \includegraphics*{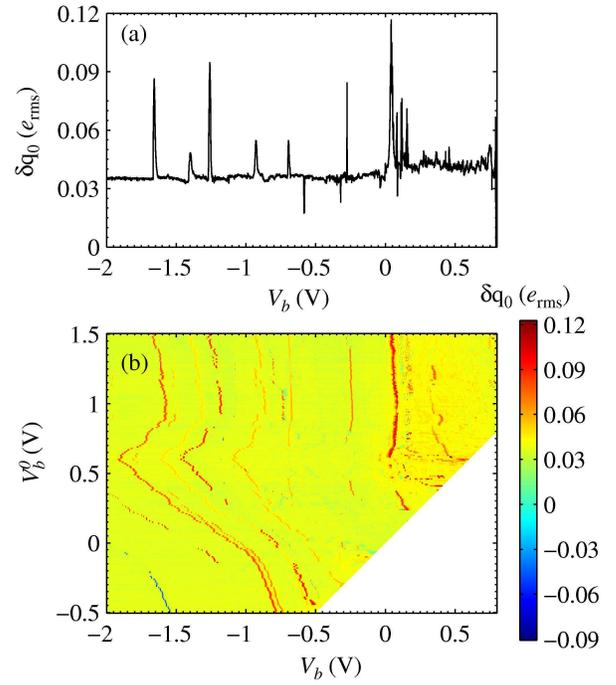}
\caption{(a) Polarizability $\delta \mathrm{q}_0$ as a function of
  bottom gate potential $V_b$. The data have been normalized to units
  of one electron on the SET island. (b) Polarizability $\delta
  \mathrm{q}_0$ (color of each point) as a function of bottom gate
  potential $V_b$ and of charge at the interface, represented by the
  initial sweep voltage $V_b^0$.}
\label{fig:polarizability}
\end{figure}

The response of the SET (as measured by the lock-in amplifier) to the
ac dither $V_\mathrm{ac}$ on the bottom gate gives a measure of the
polarizability of the intervening material. Figure\
\ref{fig:polarizability}(a) shows a typical trace of this
polarizability versus bottom gate potential, where several peaks can
be distinguished on a flat background.
The data were acquired by sweeping the bottom gate voltage while
countersweeping the top gate, thereby keeping a constant island
potential. Presumably the peaks correspond to charges moving between
distinct trapping sites when the gate potential $V_b$ brings such
sites into resonance.

\begin{figure}
  \centering \includegraphics*{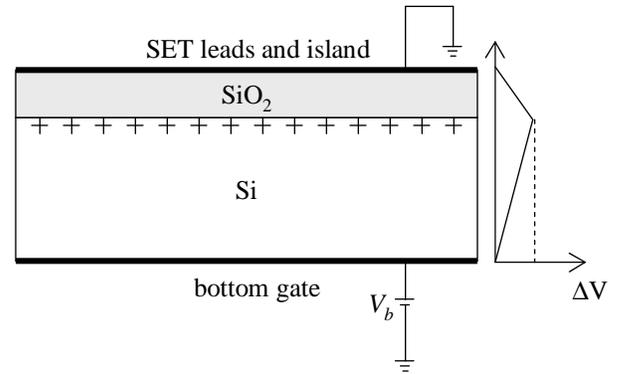}
\caption{Illustration of the effect of an increase in fixed positive
  charge at the Si/SiO$_2$ interface. The charge increases the
  electric field above the interface but decreases it below. This is
  illustrated by the diagram on the right, which shows the change in
  potential within the substrate induced by positive charge at the
  interface.}
\label{fig:electrostatics}
\end{figure}

To glean some more information about these peaks, we place some charge
at the Si/SiO$_2$ interface by biasing the bottom gate and
illuminating the sample. The actual charging process is described in
greater detail below. Such an interface charge leads to a potential
drop across the interface, changing the electrostatic configuration as
illustrated in Fig.\ \ref{fig:electrostatics}.
The resulting change in electric field within the substrate should
shift a peak along the $V_b$ axis in a direction determined by the
defect's physical location, according to the following
argument. Consider the addition of some fixed charge to the interface
that yields a change $V_b^0$ in the potential there. To restore the
electric field below the interface to its previous value requires a
corresponding increase in the potential $V_b$. However, restoring the
electric field above the interface instead necessitates an even larger
\emph{decrease} in $V_b$. If we assume that each peak occurs at a
certain electric field, then the sign of $dV_b/dV_b^0$ for a given
peak locates its associated defect above or below the Si/SiO$_2$
interface.

To create an interface charge, a negative potential (-2.0~V) is applied
to the bottom gate $V_b$ (sufficient to strongly invert the surface),
and the sample is illuminated. Referring to Fig.\ \ref{fig:set
geometry}(c), illumination creates electron-hole pairs in the Si
region. The holes flow out the bottom gate contact (to the right in
the figure), while the electrons accumulate at the Si/SiO$_2$
interface. After illumination the bottom gate is returned to a higher
potential $V_b^0$, removing some of the electrons from the interface,
although all of the electrons are depleted only when $V_b^0$ reaches
approximately 0.5~V. Then $V_b$ is swept back to -2.0~V while the
polarizability is monitored, as discussed previously. Those electrons
that have not been removed from the interface remain fixed during the
sweep.

Figure\ \ref{fig:polarizability}(b) shows a sequence of such traces,
each corresponding to a different value of $V_b^0$, now with the color
of each point representing the polarizability.  Here it is evident
that the prominent peaks indeed are shifting along the $V_b$ axis as
$V_b^0$ changes and charge is removed from the interface. Until $V_b^0
\approx 0.6$~V, the slope $dV_b^0/dV_b$ of these peaks is negative. A
negative slope rules out the substrate beneath the Si/SiO$_2$
interface as the location of the defects, a conclusion supported by
the realization that, for $V_b^0 \lesssim 0.2$~V, there exists a
two-dimensional electron gas at the interface. This electron gas
should screen the SET island completely from any charge motion in the
Si below, yet the peaks in polarizability persist in Fig.\
\ref{fig:polarizability}(b) for $V_b^0 < 0.2$~V. The absence of an
electron gas for $V_b^0 > 0.2$~V also explains the slight increase of
broadband noise here, as the SET becomes sensitive to a wider number
of noise sources below the interface.

Beyond $V_b^0 \approx 0.5$~V, all of the charge has been depleted from
the interface, and the abrupt change in $dV_b^0/dV_b$ near $V_b^0 =
0.6$~V agrees with this value. Under such conditions we expect the
peak positions to become independent of $V_b^0$, so that the slopes
$dV_b^0/dV_b$ should become nearly infinite. This is indeed the case
for $V_b^0 > 0.9$~V, but we note that several of the peaks acquire a
positive slope for $0.6\ \mathrm{V} < V_b^0 < 0.9\ \mathrm{V}$. This
positive slope is not presently understood, although we suspect that
it could be due to slow charge relaxation within the Si. For example,
successive traces of $\delta \mathrm{q}_0(V_b)$ separated by a few
minutes often show drifting of the peak positions, and in general the
peak positions are stable only when care is taken to use identical
initial conditions for successive traces. Positive $dV_b^0/dV_b$ is
consistent with a slow drift in time of the peaks toward larger values
of $V_b$, because a given value for $V_b$ occurs at later times
relative to the start of the trace with increasing $V_b^0$.

To estimate the signal expected for a single charge moving in the
substrate, we did a finite element analysis of our SET geometry using
the FEMLAB physics modeling package. The analysis indicates that the
maximum signal due to a single charge moving anywhere in the Si would
be at most 0.09~$e$ rms, yet in Fig.\ \ref{fig:polarizability} there
are at least four peaks with magnitudes near this value. For charge
motion within the SiO$_2$, our analysis gives a sensitivity gradient of
0.037~$e$/nm. A 0.09~$e$ rms signal therefore corresponds to a charge
moving about 7~nm within the oxide, a distance we find
implausible. Thus the signal magnitude indicates that the defects we
observe lie not only above the Si but very near or above the oxide
surface itself.

Defects within the tunnel barriers could easily lead to large charge
offset signals. However, these defects are located between a lead held
at constant potential and the SET island, also held at constant
potential by the top gate and feedback circuit. The strong dependence
of the observed peaks on the bottom gate voltage therefore rules out the
tunnel barriers as their origin.

We have observed similar anomalies in seven SETs of this design, as
well as in SETs fabricated on undoped Si substrates using surface
gates. Thus they would seem to be typical features in Al-AlO$_x$-Al
SETs on oxidized Si substrates. We have excluded the Si, the SiO$_2$,
and the tunnel barriers as the location of the defects associated with
these peaks. They must therefore reside very near the substrate
surface.  One possibility is that the peaks are associated with Al
grains near the SET island. Many such isolated grains are clearly
visible in magnified SEM images of our devices, although this
hypothesis awaits further studies. Another group has observed
excellent charge offset stability in Si-based SETs that should lack
such grains,\cite{Zimmerman2} an observation that may well support our
conjecture. Future experiments to eliminate these grains are
compelling in light of the importance of metal SETs to quantum
information processing, nanoelectronics, and fundamental sensing.

This work was supported by the National Security Agency.
\bibliography{bibliography}
\end{document}